\documentclass[11pt]{article}

\usepackage{amsmath}
\usepackage{color}
\usepackage{amssymb}
\usepackage{graphicx}
\usepackage{subfigure}
\usepackage{rotating}
\usepackage{setspace}
\usepackage{overcite}

\begin{document}

\title{\bf{Process Flow Diagram of an Ammonia Plant as a Complex Network}}

\author{{\bf{Zhi-Qiang Jiang, Wei-Xing Zhou\footnote{Correspondence concerning this article should be addressed to W.-X.
Zhou at wxzhou@moho.ess.ucla.edu}, Bing Xu, and Wei-Kang Yuan}}
\\State Key Laboratory of Chemical Engineering and Research Center of Systems Engineering,\\
East China University of Science and Technology, Shanghai 200237,
China}

\maketitle

\begin{abstract}
{\textit{Complex networks have attracted increasing interests in
almost all disciplines of natural and social sciences. However, few
efforts have been afforded in the field of chemical engineering. We
present in this work an example of complex technological network,
investigating the process flow of an ammonia plant (AP). We show
that the AP network is a small-world network with scale-free
distribution of degrees. Adopting Newman's maximum modularity
algorithm for the detection of communities in complex networks,
evident modular structures are identified in the AP network, which
stem from the modular sections in chemical plants. In addition, we
find that the resultant AP tree exhibits excellent allometric
scaling.}}
\end{abstract}


\section*{Introduction}
\label{sec:intro}

Complex systems are ubiquitous in natural and social sciences. The
behavior of complex system as a whole is usually richer than the sum
of its parts and it is lost if one looks at the constituents
separately. Complex systems evolve in a self-adaptive manner and
self-organize to form emergent behaviors due to the interactions
among the constituents of a complex system at the microscopic level.
The study of complexity has been witnessed in almost all disciplines
of social and natural sciences (see, for instance, the special issue
of Nature on this topic in 2001 \cite{Ziemelis-2001-Nature}).
However, engineers seem a little bit indifferent as if engineering
is at the edge of the science of complexity. Ottino argues that
``engineering should be at the centre of these developments, and
contribute to the development of new theory and tools''
\cite{Ottino-2004-Nature} and chemical engineering is facing new
opportunities \cite{Ottino-2005-AIChE}.

The topological aspects of complex systems can be modelled by
complex networks, where the constituents are viewed as vertices or
nodes and an edge is drawn between two vertices if their associated
constituents interact in certain manners. In recent years, complex
networks have attracted extensive interests, covering biological
systems, social systems, information systems, and technological
systems
\cite{Albert-Barabasi-2002-RMP,Newman-2003-SIAMR,Dorogovtsev-Mendes-2003}.
Complex networks possess many interesting properties. Most complex
networks exhibit small-world traits
\cite{Watts-Strogatz-1998-Nature} and are scale free where the
distributions of degrees have power-law tails
\cite{Barabasi-Albert-1999-Science}. In addition, many real networks
have modular structures or communities \cite{Newman-2004-EPJB}. The
fourth intriguing feature of some real networks reported recently is
the self-similarity \cite{Song-Havlin-Makse-2005-Nature}. The
studies of complex networks have extensively broadened and deepened
our understanding of complex systems.

In the field of chemical engineering, chemical reactions and
transports of mass, energy and momentum have been the traditional
domains for about five decades, where the topological properties are
of less concerns. Amaral and Ottino have considered two examples for
which the way constituents of the system are linked determines
transport and the dynamics of the system, that is, food webs and
cellular networks \cite{Amaral-Ottino-2004-CES}. In this paper, we
present an example of complex technological network in traditional
chemical engineering, studying the topological properties of the
process flow of an ammonia plant.

\begin{figure}[htb]
\begin{center}
\includegraphics[width=18cm]{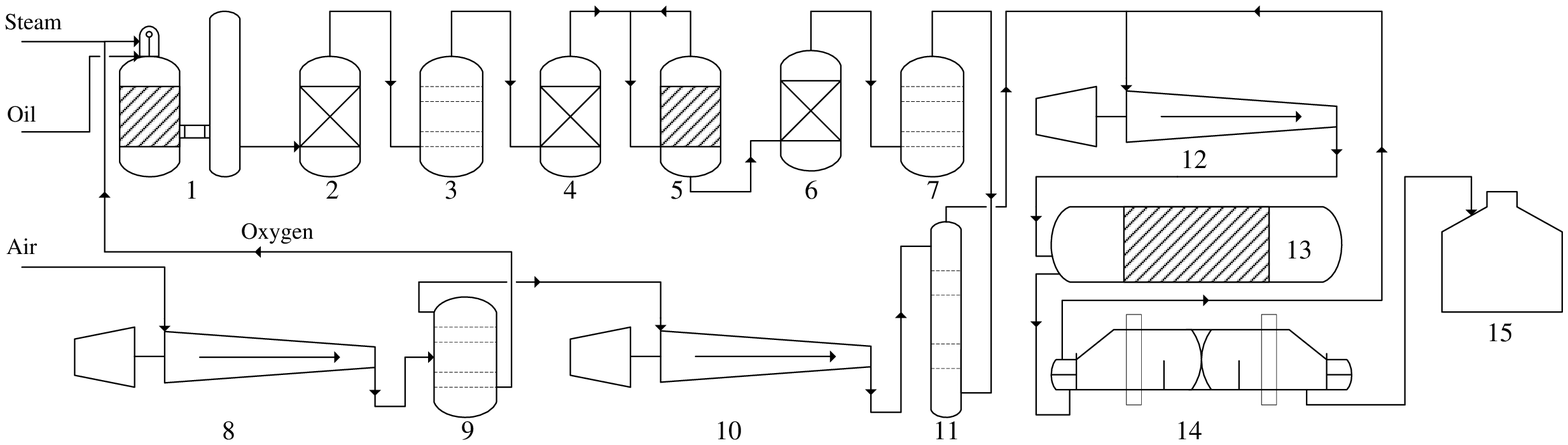}
\end{center}
\caption{Ammonia plant process flow diagram with the major
equipment. 1~--~gasification reactor, 2~--~carbon scrubber,
3~--~H$_2$S absorber, 4~--~humidifier, 5~--~shift converter,
6~--~dehumidifier, 7~--~CO$_2$absorber, 8~--~air compressor,
9~--~pressure column, 10~--~nitrogen compressor, 11~--~nitrogen wash
column, 12~--~synthesis gas compressor, 13~--~ammonia converter,
14~--~unitized chiller, 15~--~liquid ammonia tank. }
\label{Fig:Digram}
\end{figure}

The network studied here is abstracted from the process flow diagram
of the Ammonia Plant of Jiujiang Chemical Fertilizer Plant (Jiangxi
Province, China). The scale of the plant is 1000MT/D. The process
flow diagram with the major equipment is shown in
Fig.~\ref{Fig:Digram}. In the construction of the Ammonia Plant
network (AP network), towers, reactors, pumps, heat exchangers, and
connection points of convergence and bifurcation of pipes are
regarded as vertices. Only the equipments and pipes carrying raw
materials, by-products, and products are considered in the
construction of network. The utility flows are not included in the
network. The pipes connecting the vertices are treated as edges. The
AP network constructed has 505 vertices and 759 edges.

\section*{AP network exhibits small-world effect}
\label{s1:Small-world}

The average minimum path length is among the most studied quantity
in complex networks
\cite{Albert-Barabasi-2002-RMP,Newman-2003-SIAMR,Dorogovtsev-Mendes-2003}.
When regarding the AP network as an undirected network, we compute
the average minimum path length $\langle{l}\rangle=7.76$ with a
standard deviation $\sigma_l=2.65$. We find that the distribution of
$l$ is Gaussian. The skewness is 0.17 and the kurtosis excess is
0.01, which is close to the theoretical value 0 of a Gaussian
distribution. The average minimum path length and its fluctuation
can also be estimated by a Gaussian fit to the data, which presents
$\langle{l}\rangle =7.85$ and $\sigma_l =2.74$.

In most small-world networks, the average minimum path length is
somewhat larger than that for a random graph
\cite{Watts-Strogatz-1998-Nature}. It is interesting to compare the
average minimum path length of the real ammonia plant network with
that of model networks. The null model is the maximally random
networks with the same number of nodes and the same degree sequence
as the real network. There are several methods for the generation of
random graphs with prescribed degree sequences and the chain
switching method gives accurate results with acceptable
computational time
\cite{Milo-Kashtan-Itzkovitz-Newman-Alon-2004-XXX}, which was used
in the detection of rich-club structure
\cite{Colizza-Flammini-Serrano-Vespignani-2006-NP} and is the very
null model in the statistical tests of network topological
properties \cite{Amaral-Guimera-2006-NP}. Adopting the chain
switching method, we have generated 12400 random networks. The
average minimum path length $l_{\rm{rand}}$ of each model network is
calculated. It is found that $l_{\rm{rand}}= 5.90 \pm 0.07$. What is
striking is that the maximum of $l_{\rm{rand}}$ is 6.15, much
smaller than $\langle{l}\rangle=7.76$.

The clustering coefficient $C_i$ of vertex $i$ is a measure of the
cluster structure indicating how much the adjacent vertices of the
adjacent vertices of $i$ are adjacent vertices of $i$.
Mathematically, $C_i$ is defined by
\begin{equation}
 C_i = \frac{E_i}{k_i(k_i-1)/2}~,
 \label{Eq:Ci}
\end{equation}
where $E_i$ is the number of edges among the adjacent vertices of
$i$ \cite{Watts-Strogatz-1998-Nature}. The average clustering
coefficient $C=\langle C_i \rangle$ is $0.083$, which is comparable
to other technological networks \cite{Newman-2003-SIAMR}. Using the
same database of the maximally random networks, we find that
$C_{\rm{rand}}=0.0075\pm0.0036$ and the maximum clustering
coefficient of random networks is 0.025, much smaller than $C=0.083$
for the ammonia plant network. This is the evidence supporting that
the AP network is a small-world network
\cite{Watts-Strogatz-1998-Nature}.

\section*{AP network is scale-free}
\label{s1:scale-free}

The degree $k$ of a vertex of a network is the number of edges
connected to that vertex. Degree distributions of vertices are
perhaps the most frequently investigated in the literature of
complex networks
\cite{Albert-Barabasi-2002-RMP,Newman-2003-SIAMR,Dorogovtsev-Mendes-2003}.
The degree distributions of scale-free networks have fat tails
following power laws
\begin{equation}
 p(k) \sim k^{-(\mu+1)}~
 \label{Eq:pk}
\end{equation}
 Several mechanisms of scale free
distributions have been proposed, such as preferential attachment
and its variants \cite{Albert-Barabasi-2002-RMP} and fitness of
vertices
\cite{Caldarelli-Capocci-DeLosRios-Munoz-2002-PRL,Servedio-Caldarelli-Butta-2004-PRE}.
In order to estimate the probability distribution of a physical
variable empirically, several approaches are available. For a
possible power-law distribution with fat tails, cumulative
distribution or log-binning technique are usually adopted. A similar
concept to the complementary distribution, called rank-ordering
statistics \cite{Sornette-2000}, has the advantage of easy
implementation, no information loss, and being less noisy.

Consider $N$ observations of variable $k$ sampled from a
distribution whose probability density is $p(k)$. Then the
complementary distribution is $P(y > k) = \int_k^\infty p(y)
{\rm{d}}y$. We sort the $n$ observations in non-increasing order
such that $k_1\ge k_2 \ge \cdots \ge k_n \ge \cdots \ge k_n$, where
$n$ is the rank of the observation. It follows that $NP(k \ge k_n)$
is the expected number of observations larger than or equal to
$k_n$, that is,
\begin{equation}
 NP(k \ge k_n) = n~.
 \label{Eq:nPR}
\end{equation}
If the probability density of variable $k$ follows a power law that
$p(k) \sim k^{-(1+\mu)}$, then the complementary distribution $P(k)
\sim k^{-\mu}$. An intuitive relation between $k_n$ and $n$ follows
\begin{equation}
 k_n \sim n^{-1/\mu}~.
 \label{Eq:xR}
\end{equation}
A rigorous expression of (\ref{Eq:xR}) by calculating the most
probable value of $k_n$ from the probability that the $n$-th value
equals to $k_n$ gives \cite{Sornette-2000}
\begin{equation}
 k_n \sim \left(\frac{\mu N+1}{\mu n+1}\right)^{1/\mu}~.
 \label{Eq:xR0}
\end{equation}
When $\mu n\gg 1$ or equivalently $1\ll n\le N$, we retrieve
(\ref{Eq:xR}). A plot of $\ln k_n$ as a function of $\ln n$ gives a
straight line with slope $-1/\mu$ with deviations for the first a
few ranks if $k$ is distributed according to a power law of exponent
$\mu$. We note that the rank-ordering statistics is nothing but a
simple generalization of Zipf's law
\cite{Zipf-1949,Mandelbrot-1983,Sornette-2000} and has wide
applications, such as in linguistics \cite{Mandelbrot-1954-Word},
the distribution of large earthquakes
\cite{Sornette-Knopoff-Kagan-Vanneste-1996-JGR}, time-occurrences of
extreme floods \cite{Mazzarella-Rapetti-2004-JH}, to list a few.
More generally, rank-ordering statistics can be applied to
probability distributions other than power laws, such as exponential
or stretched exponential distributions
\cite{Laherrere-Sornette-1998-EPJB}, normal or log-normal
distributions \cite{Sornette-2000}, and so on.

\begin{figure}[htb]
\begin{center}
\includegraphics[width=14cm]{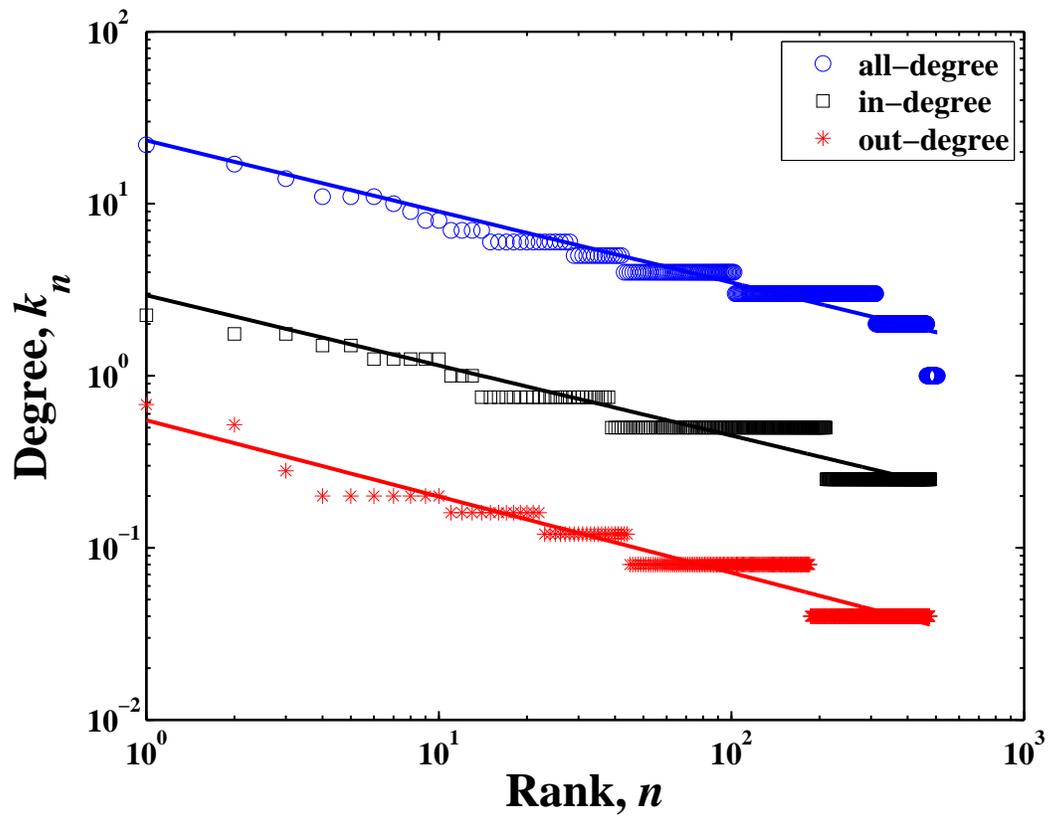}
\end{center}
\caption{Rank-ordering analysis of the in-degree, out-degree, and
all-degree of the AP network. We have translated vertically the
in-degree line by 4 and the out-degree line by 25 for better
presentation. The lines are the best fit of tail distribution to
(\ref{Eq:xR}).} \label{Fig:degree}
\end{figure}

In Fig.~\ref{Fig:degree} is shown the rank-ordering analysis of the
in-degree, out-degree and all-degree of the AP network in log-log
plot. We see that the AP network is scale-free. Linear regression of
$\ln k_n$ against $\ln n$ gives the following exponents: $1/\mu =
0.419\pm0.010$ for all-degree, $1/\mu = 0.407\pm0.009$ for
in-degree, and $1/\mu = 0.443\pm 0.008$ for out-degree. Therefore,
we have $\mu=2.39\pm  0.06$  for all-degree, $\mu= 2.46\pm 0.05$ for
in-degree, and $\mu=2.31\pm 0.04$ for out-degree.

\section*{Modular structure in the AP network}
\label{s1:modular}

\subsection*{\it{Brief review}}

In the recent years, much attention has been attracted to the
modular clusters or community structures of real networks, such as
metabolic networks
\cite{Ravasz-Somera-Mongru-Oltvai-Barabasi-2002-Science,Guimera-Amaral-2005-Nature},
food webs
\cite{Girvan-Newman-2002-PNAS,Krause-Frank-Mason-Ulanowicz-Taylor-2003-Nature},
social networks
\cite{Girvan-Newman-2002-PNAS,Guimera-Danon-DiazGuilera-Giralt-Arenas-2003-PRE,Gleiser-Danon-2003-ACS,Newman-2004-PNAS},
to list a few.

There are rigorous definitions for community. A strong community is
defined as a subgraph of the network requiring more connections
within each community than with the rest of the network, while in a
weak community the total number of connections of within-community
vertices is larger than the number of connections of the vertices in
the community with the rest of the network
\cite{Radicchi-Castellano-Cecconi-Loreto-Parisi-2004-PNAS,Castellano-Cecconi-Loreto-Parisi-Radicchi-2004-EPJB}.
However, in most cases in the literature, community is only fuzzily
defined in the sense that the connections within communities are
denser than between communities.

Different types of algorithms have been developed for the detection
of communities \cite{Newman-2004-EPJB}. Sokal and Michener proposed
the average-linkage method \cite{Sokal-Michener-1958-UKSC}, which
was extended to the hierarchical clustering algorithm later
\cite{Eisen-Spellman-Brown-Botstein-1998-PNAS}. In 1995, Frank
developed a method for direct identification of non-overlapping
subgroups \cite{Frank-1995-SN}, which was applied to detect
compartments in food webs
\cite{Krause-Frank-Mason-Ulanowicz-Taylor-2003-Nature}. Girvan and
Newman proposed a divisive algorithm that uses edge betweenness
centrality to identify the boundaries of communities
\cite{Girvan-Newman-2002-PNAS,Newman-Girvan-2004-PRE}, which is now
widely known as GN algorithm. Based on the concept of network random
walking, Zhou used dissimilarity index to delimit the boundaries of
communities, which was reported to outperform the algorithm based on
the concept of edge betweenness centrality
\cite{Zhou-2003a-PRE,Zhou-2003b-PRE}. An alternative divisive
algorithm of Radicchi {\it{et al.}} is based on the edge clustering
coefficient, related to the number of cycles that include a certain
edge
\cite{Radicchi-Castellano-Cecconi-Loreto-Parisi-2004-PNAS,Castellano-Cecconi-Loreto-Parisi-Radicchi-2004-EPJB}.
Another well-known algorithm is Newman's maximum modularity
algorithm, which is a type of agglomerative algorithm
\cite{Newman-2004-PRE,Clauset-Newman-Moore-2004-PRE}.

Many other algorithms have been presented, for instance, the
Kernighan-Lin algorim \cite{Kernighan-Lin-1970-BSTJ}, the spectral
method which takes into account weights and link orientations and
its improvement
\cite{Capocci-Servedio-Caldarelli-Colaiori-2004-LNCS,Donetti-Munoz-2005},
the resistor network approach which concerns the voltage drops
\cite{Wu-Huberman-2004-EPJB}, the information centrality algorithm
that consists in finding and removing iteratively the edge with the
highest information centrality
\cite{Fortunato-Latora-Marchiori-2004-PRE}, a fast community
detection algorithm based on a $q$-state Potts model
\cite{Reichardt-Bornholdt-2004-PRL}, an aggregation algorithm for
finding communities of related genes
\cite{Wilkinson-Huberman-2004-PNAS}, the maximum modularity
algorithm incorporated with simulated annealing
\cite{Guimera-Amaral-2005-Nature}, the agent-based algorithm
\cite{Young-Sager-Csardi-Haga-2004-XXX}, the shell algorithm
\cite{Bagrow-Bollt-2005-PRE}, and the algorithm based on random
Ising model and maximum flow \cite{Son-Jeong-Noh-2006-EPJB}.

\subsection*{\it{Community structure of the AP network}}

\begin{figure}[tb]
\begin{center}
\includegraphics[width=14cm,height=14cm]{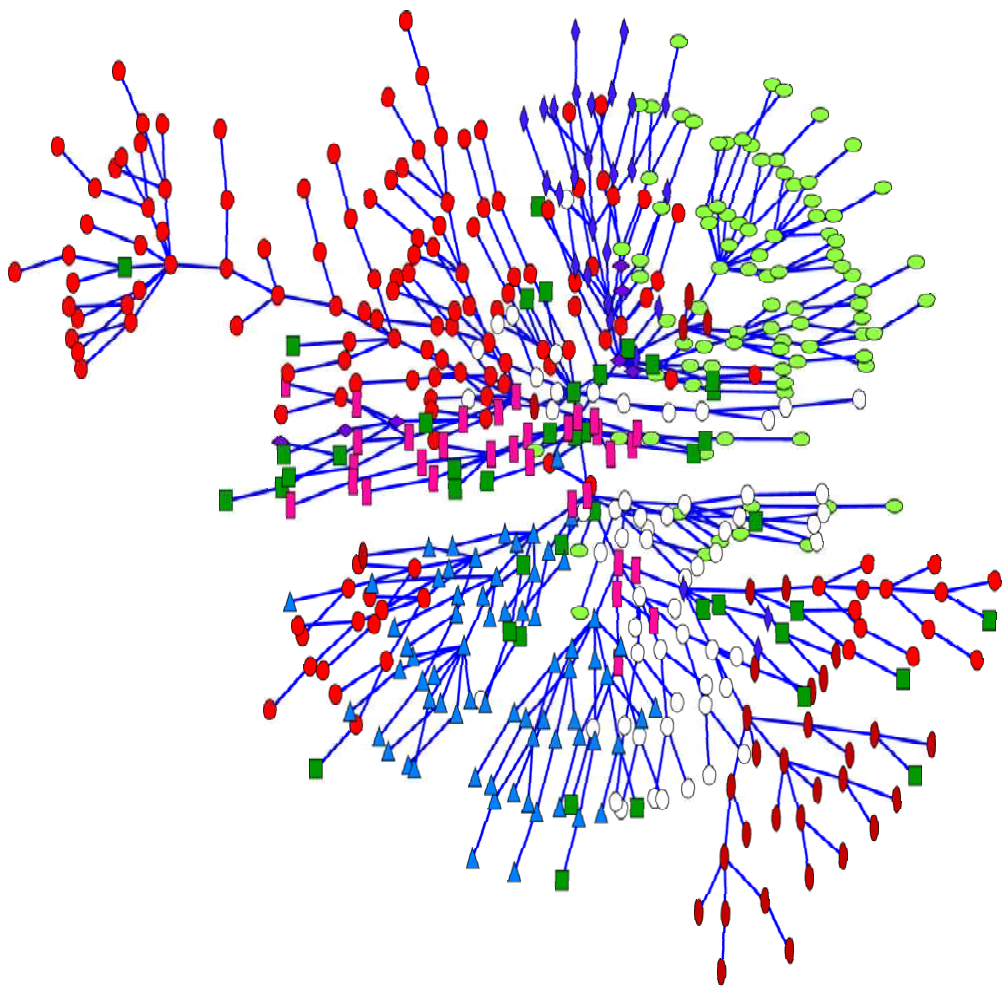}
\end{center}
\caption{(Colour online) Modular structure of the AP network. The
shapes of the vertices represent different sections of the process
flow of the Ammonia Plant. This figure was produced with Pajek
\cite{deNooy-Mrvar-Batagelj-2005}.}\label{Fig:Community}
\end{figure}

We apply Newman's maximum modularity algorithm
\cite{Newman-2004-PRE,Clauset-Newman-Moore-2004-PRE} to study the
community structure of the AP network. The resultant AP tree is
illustrated in Fig.~\ref{Fig:Community}, which is not in the form of
dendrogram. The shapes of the vertices represent different sections
of the process flow of the AP: SGP section-oil (solid circles),
rectisol section-oil (horizontal ellipses), CO-shift section
(vertical ellipses), synthesis \& refrig. section (open circles),
air separation section (triangles), nitrogen washing section
(vertical diamonds), steam superheater unit (horizontal diamonds),
ammonia storage \& tank yard (rectangles), and equipments of waste
treatment (squares). The maximum value of the modularity is
$\bf{Q}=0.794$, which is among the largest peak modularity values
reported for different networks (if not the largest) and thus
indicates a very strong community structure in the investigated
network.

It has been found that random graphs and scale-free networks have
modularity with analytic
expressions,\cite{Guimera-Sales-Pardo-Amaral-2004-PRE} which allows
us to check if the modularity observed in the AP network is
mathematically significant or not. Since the modularity of a
scale-free network with $S=505$ nodes and connectivity $m=749/505$
is
\begin{equation}\label{Eq:Modularity:SF}
    Q_{\rm{SF}} =
    \left(1-\frac{2}{\sqrt{S}}\right)\left(a+\frac{1-a}{m}\right)
    = 0.6773~,
\end{equation}
which is again much smaller than $Q=0.794$, showing that the
modularity of the AP network is significant. Note the
$a=0.165\pm0.009$.\cite{Guimera-Sales-Pardo-Amaral-2004-PRE} For an
Erd{\"{o}}s-Renyi random graph with $S=505$ nodes and connection
probability $p=749/(505\times504/2)=0.0059$, the maximal modularity
is
\begin{equation}\label{Eq:Modularity:ER}
    Q_{\rm{ER}} =
    \left(1-\frac{2}{\sqrt{S}}\right)\left(1-\frac{2}{pS}\right)^{2/3}
    = 0.6995~.
\end{equation}
The fact that $Q$ is greater than $Q_{\rm{ER}}$ indicates that the
modular structure extracted from the AP network could not be
attributed to the fluctuation of random graphs and is thus still
very significant.

Alternatively, we can use the same null model which employs the
chain switching algorithm to generate maximally random networks with
the same degree sequence of the AP network. we find that
$Q_{\rm{rand}}=0.440\pm0.009$ and the maximum of the modularity of
model networks is 0.469, which is much smaller than $Q=0.794$. This
test provide further evidence that the modular structure in the AP
network is statistically significant.

The modular structures of chemical plant networks do not come out as
a surprise. In a chemical plant, raw materials are fed into the
process flow network and react from one section to another
successively, although there are feedbacks from later sections. In
general, flows are denser within a workshop section than between
sections. Therefore, a section is naturally a community. In
Fig.~\ref{Fig:Community}, most of the vertices in a given section
are recognized to be members of a same community. The vertices of
the storage and tank yard (rectangles) are the most dispensed in
Fig.~\ref{Fig:Community}. This is expected since these tanks are
linked from and to different sections in the process, which shows
the power of Newman's maximum modularity algorithm for community
detection.

\subsection*{\it{Allometric scaling of the AP tree}}

The network shown in Fig.~\ref{Fig:Community} is actually a tree.
Trees exhibit intriguing intrinsic properties other than non-tree
networks, among which is the allometric scaling. Allometric scaling
laws are ubiquitous in networking systems such as metabolism of
organisms and ecosystems river networks, food webs, and so forth
\cite{West-Brown-Enquist-1997-Science,Enquist-Brown-West-1998-Nature,West-Brown-Enquist-1999-Science,Enquist-West-Charnov-Brown-1999-Nature,Banavar-Maritan-Rinaldo-1999-Nature,Enquist-Economo-Huxman-Allen-Ignace-Gillooly-2003-Nature,Garlaschelli-Caldarelli-Pietronero-2003-Nature}.
The original model of the allometric scaling on a spanning tree was
developed by Banavar, Maritan, and Rinaldo
\cite{Banavar-Maritan-Rinaldo-1999-Nature}. The spanning tree has
one root and many branches and leaves, and can be rated as directed
from root to leaves. Mathematically, each node of a tree is assigned
a number 1 and two values $A_i$ and $S_i$ are defined for each node
$i$ in a recursive manner as follows:
\begin{subequations}
\begin{equation}
A_i = \sum_j A_j + 1~,\label{Eq:A}
\end{equation}
and
\begin{equation}
 S_i = \sum_j S_j +
A_i,\label{Eq:C}
\end{equation}
\label{Eq:AC}
\end{subequations} where $j$ stands for the nodes linked {\it{from}} $i$
\cite{Banavar-Maritan-Rinaldo-1999-Nature}. In a food web, $i$ is
the prey and $j$'s are its predators (thus the nutrition flows from
$i$ to $j$'s). The allometric scaling relation is then highlighted
by the power law relation between $S_i$ and $A_i$:
\begin{equation}
 S \sim A^{\eta}~.
 \label{Eq:AC:eta}
\end{equation}

For spanning trees extracted from transportation networks, the power
law exponent $\eta$ is a measure of transportation efficiency
\cite{Banavar-Maritan-Rinaldo-1999-Nature,Garlaschelli-Caldarelli-Pietronero-2003-Nature}.
The smaller is the value of $\eta$, the more efficient is the
transportation. Any spanning tree can range in principle between two
extremes, that is, the chain-like trees and the star-like trees. A
chain tree has one root and one leaf with no branching. Let's label
leaf vertex by 1, its father by 2, and so forth. The root is
labelled by $n$ for a chain-like tree of size $n$. The recursive
relations (\ref{Eq:AC}) become $S_i=S_{i-1}+A_i$ and $A_i=A_{i-1}+1$
with termination conditions $A_1=S_1=1$. It is easy to show that
$A_i=i$ and $S_i=i(i+1)/2$. Asymptotically, the exponent $\eta=2^-$
for chain-like trees. For star-like trees of size $n$, there are one
root and $n-1$ leaves directly connected to the root. We have
$A=S=1$ for all the leaves and $A=n$ and $S=2n-1$ for the root. It
follows approximately that $\eta=1^+$. Therefore, $1< \eta < 2$ for
all spanning trees.

We note that not all trees have such allometric scaling. Consider
for instance the classic Cayley with $n$ generations where the root
is the first generation. The $A$ and $S$ values of the vertices of
the same generation are identical. If we denote $A_i$ and $S_i$ for
the vertices of the $(n+1-i)$-th generation, the iterative equations
are $A_{i+1}=2A_i+1$ and $S_{i+1}=2S_i+A_{i+1}$, resulting in
$A_i=2^i-1$ and $S_i=(i-1)2^i+1$. This leads to
$S=[\log_2(A+1)-1]A+\log_2(A+1)$. Obviously, there is no power-law
dependence between $A$ and $S$.

\begin{figure}[htb]
\begin{center}
\includegraphics[width=14cm]{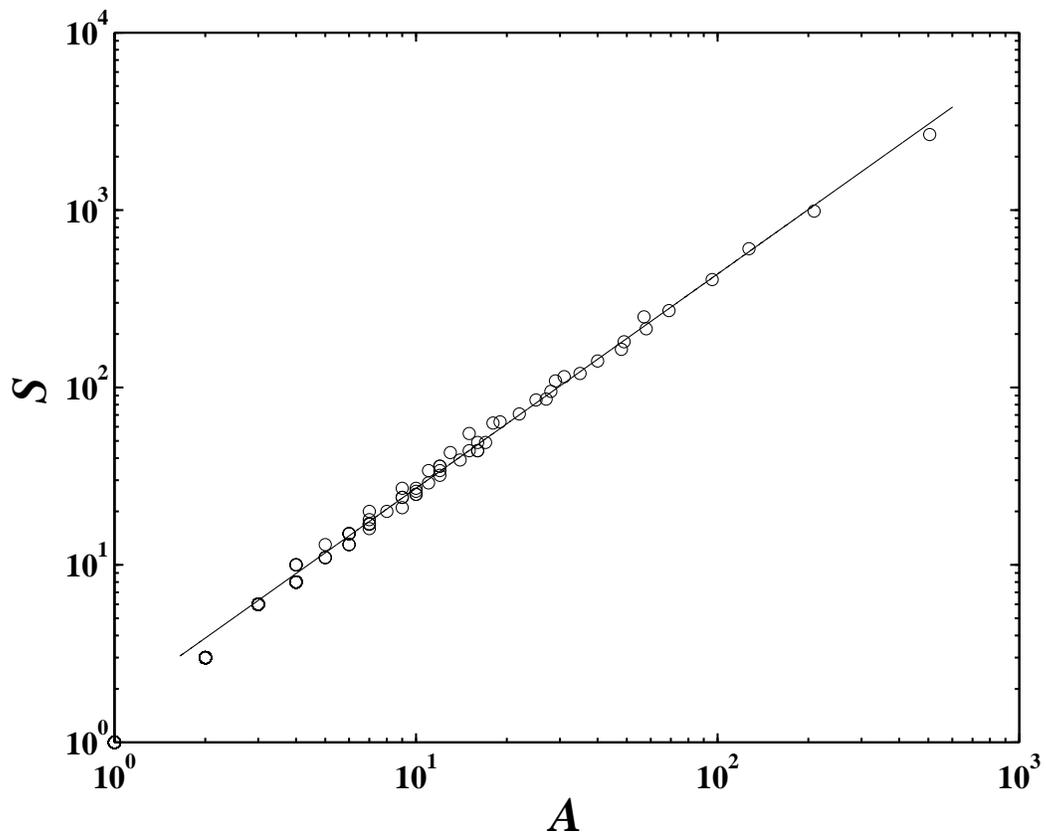}
\end{center}
\caption{Power-law scaling of $S$ against $A$. The line represents
the power-law fit to the data.} \label{Fig:AllometricScaling}
\end{figure}

We apply this framework on the AP tree. The calculated $S$ is
plotted in Fig.~\ref{Fig:AllometricScaling} as a function of $A$. A
nice power-law relation is observed between $S$ and $A$. A linear
fit of $\ln S$ against $\ln A$ give $\eta = 1.21$ with regression
coefficient 0.998. The trivial point $(A=1,S=1)$ is excluded from
the fitting \cite{Garlaschelli-Caldarelli-Pietronero-2003-Nature}.
This value of $\eta$ is slightly larger than $\eta=1.13\sim1.16$ for
food webs \cite{Garlaschelli-Caldarelli-Pietronero-2003-Nature} but
much smaller than $\eta=1.5$ for river networks
\cite{Banavar-Maritan-Rinaldo-1999-Nature}. This analysis is
relevant when the flux in the pipes and reactors are considered for
the investigation of the transportation efficiency, as an analogue
to the river network and biological network
\cite{West-Brown-Enquist-1997-Science,Banavar-Maritan-Rinaldo-1999-Nature}.

\section*{Concluding remarks}
\label{s1:conclude}

We have studied a complex technological network extracted from the
process flow of the Ammonia Plant of Jiujiang Chemical Fertilizer
Plant in Jiangxi Province of China. We have shown that the ammonia
plant network is a small-world network in the sense that its minimum
average path length $\langle{l}\rangle=7.76$ and global clustering
coefficient $C=0.083$ are respectively larger than their
counterparts $l_{\rm{rand}}=5.90\pm0.07$ and
$C_{\rm{rand}}=0.0075\pm0.0036$ of a semble of 12400 maximally
random graphs having the same degree sequences of the real AP
network. We found that the shortest path lengths between two
arbitrary vertices are distributed according to a Gaussian formula.
The distribution of degrees follows a power law with its exponent
being $\mu=2.31-2.46$, indicating that the AP network is scale-free.

We have reviewed briefly diverse existing algorithms for the
detection of community structures in complex networks, among which
Newman's maximum modularity algorithm is applied to the AP network.
The extracted modular structures have a very high modularity value
$Q=0.794$ signaling the significance of the modules, which is
confirmed by statistical tests. These modular structures are well
explained by the workshop sections of the ammonia plant. We have
constructed a spanning tree based on the community identification
procedure and found that the resultant AP tree exhibits excellent
allometric scaling with an exponent comparable to the universal
scaling exponent of food webs.

In summary, we have studied the topological properties of the AP
network from chemical engineering. More sophisticated networks can
be constructed from process flows in chemical industry. There are
still other open problems even in this small AP network, such as the
origin of the scale-free feature, what we can learn from these
topological features, robustness and sensitivity analysis on the
mass flows to find out bottlenecks in the process or figure out how
jamming of the nodes or cascade failure of the system can occur, to
list but a few. We hope that this work will attract more affords in
this direction. Further researches on complex networks containing
information of transports and reactions will unveil useful
properties and benefit the field practically and theoretically.

\bigskip\bigskip

{\bf Acknowledgment}

The authors thank gratefully Hai-Feng Liu for providing the process
flow diagram of Jiujiang Ammonia Plant. This work was partially
supported by National Basic Research Program of China (No.
2004CB217703), the Project Sponsored by the Scientific Research
Foundation for the Returned Overseas Chinese Scholars, State
Education Ministry of China, and NSFC/PetroChina through a major
project on multiscale methodology (No. 20490200).

\bibliography{E:/papers/Bibliography}

\begin{thebibliography}{57}
\expandafter\ifx\csname natexlab\endcsname\relax\def\natexlab#1{#1}\fi
\expandafter\ifx\csname bibnamefont\endcsname\relax
  \def\bibnamefont#1{#1}\fi
\expandafter\ifx\csname bibfnamefont\endcsname\relax
  \def\bibfnamefont#1{#1}\fi
\expandafter\ifx\csname citenamefont\endcsname\relax
  \def\citenamefont#1{#1}\fi
\expandafter\ifx\csname url\endcsname\relax
  \def\url#1{\texttt{#1}}\fi
\expandafter\ifx\csname urlprefix\endcsname\relax\def\urlprefix{URL }\fi
\providecommand{\bibinfo}[2]{#2}
\providecommand{\eprint}[2][]{\url{#2}}

\bibitem[{\citenamefont{Ziemelis}(2001)}]{Ziemelis-2001-Nature}
\bibinfo{author}{\bibfnamefont{K.}~\bibnamefont{Ziemelis}},
  \bibinfo{journal}{Nature} \textbf{\bibinfo{volume}{410}},
  \bibinfo{pages}{241} (\bibinfo{year}{2001}).

\bibitem[{\citenamefont{Ottino}(2004)}]{Ottino-2004-Nature}
\bibinfo{author}{\bibfnamefont{J.~M.} \bibnamefont{Ottino}},
  \bibinfo{journal}{Nature} \textbf{\bibinfo{volume}{427}},
  \bibinfo{pages}{399} (\bibinfo{year}{2004}).

\bibitem[{\citenamefont{Albert and
  Barab{\'a}si}(2002)}]{Albert-Barabasi-2002-RMP}
\bibinfo{author}{\bibfnamefont{R.}~\bibnamefont{Albert}} \bibnamefont{and}
  \bibinfo{author}{\bibfnamefont{A.-L.} \bibnamefont{Barab{\'a}si}},
  \bibinfo{journal}{Rev. Mod. Phys.} \textbf{\bibinfo{volume}{74}},
  \bibinfo{pages}{47} (\bibinfo{year}{2002}).

\bibitem[{\citenamefont{Newman}(2003)}]{Newman-2003-SIAMR}
\bibinfo{author}{\bibfnamefont{M.~E.~J.} \bibnamefont{Newman}},
  \bibinfo{journal}{SIAM Rev.} \textbf{\bibinfo{volume}{45}},
  \bibinfo{pages}{167} (\bibinfo{year}{2003}).

\bibitem[{\citenamefont{Dorogovtsev and
  Mendes}(2003)}]{Dorogovtsev-Mendes-2003}
\bibinfo{author}{\bibfnamefont{S.~N.} \bibnamefont{Dorogovtsev}}
  \bibnamefont{and} \bibinfo{author}{\bibfnamefont{J.~F.~F.}
  \bibnamefont{Mendes}}, \emph{\bibinfo{title}{Evolution of Networks: From
  Biological Nets to the Internet and the WWW}} (\bibinfo{publisher}{Oxford
  University Press}, \bibinfo{address}{Oxford}, \bibinfo{year}{2003}).

\bibitem[{\citenamefont{Watts and Strogatz}(1998)}]{Watts-Strogatz-1998-Nature}
\bibinfo{author}{\bibfnamefont{D.~J.} \bibnamefont{Watts}} \bibnamefont{and}
  \bibinfo{author}{\bibfnamefont{S.~H.} \bibnamefont{Strogatz}},
  \bibinfo{journal}{Nature} \textbf{\bibinfo{volume}{393}},
  \bibinfo{pages}{440} (\bibinfo{year}{1998}).

\bibitem[{\citenamefont{Barab{\'a}si and
  Albert}(1999)}]{Barabasi-Albert-1999-Science}
\bibinfo{author}{\bibfnamefont{A.-L.} \bibnamefont{Barab{\'a}si}}
  \bibnamefont{and} \bibinfo{author}{\bibfnamefont{R.}~\bibnamefont{Albert}},
  \bibinfo{journal}{Science} \textbf{\bibinfo{volume}{286}},
  \bibinfo{pages}{509} (\bibinfo{year}{1999}).

\bibitem[{\citenamefont{Newman}(2004{\natexlab{a}})}]{Newman-2004-EPJB}
\bibinfo{author}{\bibfnamefont{M.~E.~J.} \bibnamefont{Newman}},
  \bibinfo{journal}{Eur. Phys. J. B} \textbf{\bibinfo{volume}{38}},
  \bibinfo{pages}{321} (\bibinfo{year}{2004}{\natexlab{a}}).

\bibitem[{\citenamefont{Song et~al.}(2005)\citenamefont{Song, Havlin, and
  Makse}}]{Song-Havlin-Makse-2005-Nature}
\bibinfo{author}{\bibfnamefont{C.-M.} \bibnamefont{Song}},
  \bibinfo{author}{\bibfnamefont{S.}~\bibnamefont{Havlin}}, \bibnamefont{and}
  \bibinfo{author}{\bibfnamefont{H.~A.} \bibnamefont{Makse}},
  \bibinfo{journal}{Nature} \textbf{\bibinfo{volume}{433}},
  \bibinfo{pages}{392} (\bibinfo{year}{2005}).

\bibitem[{\citenamefont{Amaral and Ottino}(2004)}]{Amaral-Ottino-2004-CES}
\bibinfo{author}{\bibfnamefont{L.~A.~N.} \bibnamefont{Amaral}}
  \bibnamefont{and} \bibinfo{author}{\bibfnamefont{J.~M.}
  \bibnamefont{Ottino}}, \bibinfo{journal}{Chem. Eng. Sci.}
  \textbf{\bibinfo{volume}{59}}, \bibinfo{pages}{1653} (\bibinfo{year}{2004}).

\bibitem[{\citenamefont{Caldarelli et~al.}(2002)\citenamefont{Caldarelli,
  Capocci, De~Los~Rios, and
  Mu{\~n}oz}}]{Caldarelli-Capocci-DeLosRios-Munoz-2002-PRL}
\bibinfo{author}{\bibfnamefont{G.}~\bibnamefont{Caldarelli}},
  \bibinfo{author}{\bibfnamefont{A.}~\bibnamefont{Capocci}},
  \bibinfo{author}{\bibfnamefont{P.}~\bibnamefont{De~Los~Rios}},
  \bibnamefont{and} \bibinfo{author}{\bibfnamefont{M.~A.}
  \bibnamefont{Mu{\~n}oz}}, \bibinfo{journal}{Phys. Rev. Lett.}
  \textbf{\bibinfo{volume}{89}}, \bibinfo{pages}{258702}
  (\bibinfo{year}{2002}).

\bibitem[{\citenamefont{Servedio et~al.}(2004)\citenamefont{Servedio,
  Caldarelli, and Butt{\`a}}}]{Servedio-Caldarelli-Butta-2004-PRE}
\bibinfo{author}{\bibfnamefont{V.~D.~P.} \bibnamefont{Servedio}},
  \bibinfo{author}{\bibfnamefont{G.}~\bibnamefont{Caldarelli}},
  \bibnamefont{and}
  \bibinfo{author}{\bibfnamefont{P.}~\bibnamefont{Butt{\`a}}},
  \bibinfo{journal}{Phys. Rev. E} \textbf{\bibinfo{volume}{70}},
  \bibinfo{pages}{056126} (\bibinfo{year}{2004}).

\bibitem[{\citenamefont{Dorogovtsev et~al.}(2002)\citenamefont{Dorogovtsev,
  Goltsev, and Mendes}}]{Dorogovtsev-Goltsev-Mendes-2002-PRE}
\bibinfo{author}{\bibfnamefont{S.~N.} \bibnamefont{Dorogovtsev}},
  \bibinfo{author}{\bibfnamefont{A.~V.} \bibnamefont{Goltsev}},
  \bibnamefont{and} \bibinfo{author}{\bibfnamefont{J.~F.~F.}
  \bibnamefont{Mendes}}, \bibinfo{journal}{Phys. Rev. E}
  \textbf{\bibinfo{volume}{65}}, \bibinfo{pages}{066122}
  (\bibinfo{year}{2002}).

\bibitem[{\citenamefont{Holme and Kim}(2002)}]{Holme-Kim-2002-PRE}
\bibinfo{author}{\bibfnamefont{P.}~\bibnamefont{Holme}} \bibnamefont{and}
  \bibinfo{author}{\bibfnamefont{B.~J.} \bibnamefont{Kim}},
  \bibinfo{journal}{Phys. Rev. E} \textbf{\bibinfo{volume}{65}},
  \bibinfo{pages}{026107} (\bibinfo{year}{2002}).

\bibitem[{\citenamefont{Szab{\'o} et~al.}(2003)\citenamefont{Szab{\'o}, Alava,
  and Kert{\'e}sz}}]{Szabo-Alava-Kertesz-2003-PRE}
\bibinfo{author}{\bibfnamefont{G.}~\bibnamefont{Szab{\'o}}},
  \bibinfo{author}{\bibfnamefont{M.}~\bibnamefont{Alava}}, \bibnamefont{and}
  \bibinfo{author}{\bibfnamefont{J.}~\bibnamefont{Kert{\'e}sz}},
  \bibinfo{journal}{Phys. Rev. E} \textbf{\bibinfo{volume}{67}},
  \bibinfo{pages}{066102} (\bibinfo{year}{2003}).

\bibitem[{\citenamefont{Ravasz et~al.}(2002)\citenamefont{Ravasz, Somera,
  Mongru, Oltvai, and
  Barab{\'a}si}}]{Ravasz-Somera-Mongru-Oltvai-Barabasi-2002-Science}
\bibinfo{author}{\bibfnamefont{E.}~\bibnamefont{Ravasz}},
  \bibinfo{author}{\bibfnamefont{A.~L.} \bibnamefont{Somera}},
  \bibinfo{author}{\bibfnamefont{D.~A.} \bibnamefont{Mongru}},
  \bibinfo{author}{\bibfnamefont{A.~N.} \bibnamefont{Oltvai}},
  \bibnamefont{and} \bibinfo{author}{\bibfnamefont{A.-L.}
  \bibnamefont{Barab{\'a}si}}, \bibinfo{journal}{Science}
  \textbf{\bibinfo{volume}{297}}, \bibinfo{pages}{1551} (\bibinfo{year}{2002}).

\bibitem[{\citenamefont{Ravasz and
  Barab{\'a}si}(2003)}]{Ravasz-Barabasi-2003-PRE}
\bibinfo{author}{\bibfnamefont{E.}~\bibnamefont{Ravasz}} \bibnamefont{and}
  \bibinfo{author}{\bibfnamefont{A.-L.} \bibnamefont{Barab{\'a}si}},
  \bibinfo{journal}{Phys. Rev. E} \textbf{\bibinfo{volume}{67}},
  \bibinfo{pages}{026112} (\bibinfo{year}{2003}).

\bibitem[{\citenamefont{V{\'a}zquez}(2003)}]{Vazquez-2003-PRE}
\bibinfo{author}{\bibfnamefont{A.}~\bibnamefont{V{\'a}zquez}},
  \bibinfo{journal}{Phys. Rev. E} \textbf{\bibinfo{volume}{67}},
  \bibinfo{pages}{056104} (\bibinfo{year}{2003}).

\bibitem[{\citenamefont{Albert et~al.}(2000)\citenamefont{Albert, Jeong, and
  Barab{\'a}si}}]{Albert-Jeong-barabasi-2000-Nature}
\bibinfo{author}{\bibfnamefont{R.}~\bibnamefont{Albert}},
  \bibinfo{author}{\bibfnamefont{H.}~\bibnamefont{Jeong}}, \bibnamefont{and}
  \bibinfo{author}{\bibfnamefont{A.-L.} \bibnamefont{Barab{\'a}si}},
  \bibinfo{journal}{Nature} \textbf{\bibinfo{volume}{406}},
  \bibinfo{pages}{378} (\bibinfo{year}{2000}).

\bibitem[{\citenamefont{Kim}(2004)}]{Kim-2004-PRL}
\bibinfo{author}{\bibfnamefont{B.~J.} \bibnamefont{Kim}},
  \bibinfo{journal}{Phys. Rev. Lett.} \textbf{\bibinfo{volume}{93}},
  \bibinfo{pages}{168701} (\bibinfo{year}{2004}).

\bibitem[{\citenamefont{Strogatz}(2005)}]{Strogatz-2005-Nature}
\bibinfo{author}{\bibfnamefont{S.~H.} \bibnamefont{Strogatz}},
  \bibinfo{journal}{Nature} \textbf{\bibinfo{volume}{433}},
  \bibinfo{pages}{365} (\bibinfo{year}{2005}).

\bibitem[{\citenamefont{Zhou et~al.}(2005)\citenamefont{Zhou, Sornette, Hill,
  and Dunbar}}]{Zhou-Sornette-Hill-Dunbar-2005-PRSB}
\bibinfo{author}{\bibfnamefont{W.-X.} \bibnamefont{Zhou}},
  \bibinfo{author}{\bibfnamefont{D.}~\bibnamefont{Sornette}},
  \bibinfo{author}{\bibfnamefont{R.~A.} \bibnamefont{Hill}}, \bibnamefont{and}
  \bibinfo{author}{\bibfnamefont{R.~I.~M.} \bibnamefont{Dunbar}},
  \bibinfo{journal}{Proc. Royal Soc. B} \textbf{\bibinfo{volume}{272}},
  \bibinfo{pages}{439} (\bibinfo{year}{2005}).

\bibitem[{\citenamefont{Meakin}(1998)}]{Meakin-1998}
\bibinfo{author}{\bibfnamefont{P.}~\bibnamefont{Meakin}},
  \emph{\bibinfo{title}{Fractals, scaling and growth far from equilibrium}}
  (\bibinfo{publisher}{Cambridge University Press}, \bibinfo{address}{London},
  \bibinfo{year}{1998}).

\bibitem[{\citenamefont{Guimer{\`a} and
  Amaral}(2005)}]{Guimera-Amaral-2005-Nature}
\bibinfo{author}{\bibfnamefont{R.}~\bibnamefont{Guimer{\`a}}} \bibnamefont{and}
  \bibinfo{author}{\bibfnamefont{L.~A.~N.} \bibnamefont{Amaral}},
  \bibinfo{journal}{Nature} \textbf{\bibinfo{volume}{433}},
  \bibinfo{pages}{895} (\bibinfo{year}{2005}).

\bibitem[{\citenamefont{Girvan and Newman}(2002)}]{Girvan-Newman-2002-PNAS}
\bibinfo{author}{\bibfnamefont{M.}~\bibnamefont{Girvan}} \bibnamefont{and}
  \bibinfo{author}{\bibfnamefont{M.~E.~J.} \bibnamefont{Newman}},
  \bibinfo{journal}{Proc. Natl. Acad. Sci. USA} \textbf{\bibinfo{volume}{99}},
  \bibinfo{pages}{7821} (\bibinfo{year}{2002}).

\bibitem[{\citenamefont{Krause et~al.}(2003)\citenamefont{Krause, Frank, Mason,
  Ulanowicz, and Taylor}}]{Krause-Frank-Mason-Ulanowicz-Taylor-2003-Nature}
\bibinfo{author}{\bibfnamefont{A.~E.} \bibnamefont{Krause}},
  \bibinfo{author}{\bibfnamefont{K.~A.} \bibnamefont{Frank}},
  \bibinfo{author}{\bibfnamefont{D.~M.} \bibnamefont{Mason}},
  \bibinfo{author}{\bibfnamefont{R.~E.} \bibnamefont{Ulanowicz}},
  \bibnamefont{and} \bibinfo{author}{\bibfnamefont{W.~W.}
  \bibnamefont{Taylor}}, \bibinfo{journal}{Nature}
  \textbf{\bibinfo{volume}{426}}, \bibinfo{pages}{282} (\bibinfo{year}{2003}).

\bibitem[{\citenamefont{Guimer{\`a} et~al.}(2003)\citenamefont{Guimer{\`a},
  Danon, D{\'i}az-Guilera, Giralt, and
  Arenas}}]{Guimera-Danon-DiazGuilera-Giralt-Arenas-2003-PRE}
\bibinfo{author}{\bibfnamefont{R.}~\bibnamefont{Guimer{\`a}}},
  \bibinfo{author}{\bibfnamefont{L.}~\bibnamefont{Danon}},
  \bibinfo{author}{\bibfnamefont{A.}~\bibnamefont{D{\'i}az-Guilera}},
  \bibinfo{author}{\bibfnamefont{F.}~\bibnamefont{Giralt}}, \bibnamefont{and}
  \bibinfo{author}{\bibfnamefont{A.}~\bibnamefont{Arenas}},
  \bibinfo{journal}{Phys. Rev. E} \textbf{\bibinfo{volume}{68}},
  \bibinfo{pages}{065103} (\bibinfo{year}{2003}).

\bibitem[{\citenamefont{Gleiser and Danon}(2003)}]{Gleiser-Danon-2003-ACS}
\bibinfo{author}{\bibfnamefont{P.~M.} \bibnamefont{Gleiser}} \bibnamefont{and}
  \bibinfo{author}{\bibfnamefont{L.}~\bibnamefont{Danon}},
  \bibinfo{journal}{Adv. Complex Systems} \textbf{\bibinfo{volume}{6}},
  \bibinfo{pages}{565} (\bibinfo{year}{2003}).

\bibitem[{\citenamefont{Newman}(2004{\natexlab{b}})}]{Newman-2004-PNAS}
\bibinfo{author}{\bibfnamefont{M.~E.~J.} \bibnamefont{Newman}},
  \bibinfo{journal}{Proc. Natl. Acad. Sci. USA} \textbf{\bibinfo{volume}{101}},
  \bibinfo{pages}{5200} (\bibinfo{year}{2004}{\natexlab{b}}).

\bibitem[{\citenamefont{Radicchi et~al.}(2004)\citenamefont{Radicchi,
  Castellano, Cecconi, Loreto, and
  Parisi}}]{Radicchi-Castellano-Cecconi-Loreto-Parisi-2004-PNAS}
\bibinfo{author}{\bibfnamefont{F.}~\bibnamefont{Radicchi}},
  \bibinfo{author}{\bibfnamefont{C.}~\bibnamefont{Castellano}},
  \bibinfo{author}{\bibfnamefont{F.}~\bibnamefont{Cecconi}},
  \bibinfo{author}{\bibfnamefont{V.}~\bibnamefont{Loreto}}, \bibnamefont{and}
  \bibinfo{author}{\bibfnamefont{D.}~\bibnamefont{Parisi}},
  \bibinfo{journal}{Proc. Natl. Acad. Sci. USA} \textbf{\bibinfo{volume}{101}},
  \bibinfo{pages}{2658} (\bibinfo{year}{2004}).

\bibitem[{\citenamefont{Castellano et~al.}(2004)\citenamefont{Castellano,
  Cecconi, Loreto, Parisi, and
  F}}]{Castellano-Cecconi-Loreto-Parisi-Radicchi-2004-EPJB}
\bibinfo{author}{\bibfnamefont{C.}~\bibnamefont{Castellano}},
  \bibinfo{author}{\bibfnamefont{F.}~\bibnamefont{Cecconi}},
  \bibinfo{author}{\bibfnamefont{V.}~\bibnamefont{Loreto}},
  \bibinfo{author}{\bibfnamefont{D.}~\bibnamefont{Parisi}}, \bibnamefont{and}
  \bibinfo{author}{\bibfnamefont{R.}~\bibnamefont{F}}, \bibinfo{journal}{Eur.
  Phys. J. B} \textbf{\bibinfo{volume}{38}}, \bibinfo{pages}{311}
  (\bibinfo{year}{2004}).

\bibitem[{\citenamefont{Sokal and Michener}(1958)}]{Sokal-Michener-1958-UKSC}
\bibinfo{author}{\bibfnamefont{R.~R.} \bibnamefont{Sokal}} \bibnamefont{and}
  \bibinfo{author}{\bibfnamefont{C.~D.} \bibnamefont{Michener}},
  \bibinfo{journal}{Univ. Kans. Sci. Bull.} \textbf{\bibinfo{volume}{38}},
  \bibinfo{pages}{1409} (\bibinfo{year}{1958}).

\bibitem[{\citenamefont{Eisen et~al.}(1998)\citenamefont{Eisen, Spellman,
  Brown, and Botstein}}]{Eisen-Spellman-Brown-Botstein-1998-PNAS}
\bibinfo{author}{\bibfnamefont{M.~B.} \bibnamefont{Eisen}},
  \bibinfo{author}{\bibfnamefont{P.~T.} \bibnamefont{Spellman}},
  \bibinfo{author}{\bibfnamefont{P.~O.} \bibnamefont{Brown}}, \bibnamefont{and}
  \bibinfo{author}{\bibfnamefont{D.}~\bibnamefont{Botstein}},
  \bibinfo{journal}{Proc. Natl. Acad. Sci. USA} \textbf{\bibinfo{volume}{85}},
  \bibinfo{pages}{14863} (\bibinfo{year}{1998}).

\bibitem[{\citenamefont{Frank}(1995)}]{Frank-1995-SN}
\bibinfo{author}{\bibfnamefont{K.~A.} \bibnamefont{Frank}},
  \bibinfo{journal}{Soc. Networks} \textbf{\bibinfo{volume}{17}},
  \bibinfo{pages}{27} (\bibinfo{year}{1995}).

\bibitem[{\citenamefont{Newman and Girvan}(2004)}]{Newman-Girvan-2004-PRE}
\bibinfo{author}{\bibfnamefont{M.~E.~J.} \bibnamefont{Newman}}
  \bibnamefont{and} \bibinfo{author}{\bibfnamefont{M.}~\bibnamefont{Girvan}},
  \bibinfo{journal}{Phys. Rev. E} \textbf{\bibinfo{volume}{69}},
  \bibinfo{pages}{026113} (\bibinfo{year}{2004}).

\bibitem[{\citenamefont{Zhou}(2003{\natexlab{a}})}]{Zhou-2003a-PRE}
\bibinfo{author}{\bibfnamefont{H.~J.} \bibnamefont{Zhou}},
  \bibinfo{journal}{Phys. Rev. E} \textbf{\bibinfo{volume}{67}},
  \bibinfo{pages}{041908} (\bibinfo{year}{2003}{\natexlab{a}}).

\bibitem[{\citenamefont{Zhou}(2003{\natexlab{b}})}]{Zhou-2003b-PRE}
\bibinfo{author}{\bibfnamefont{H.~J.} \bibnamefont{Zhou}},
  \bibinfo{journal}{Phys. Rev. E} \textbf{\bibinfo{volume}{67}},
  \bibinfo{pages}{061901} (\bibinfo{year}{2003}{\natexlab{b}}).

\bibitem[{\citenamefont{Newman}(2004{\natexlab{c}})}]{Newman-2004-PRE}
\bibinfo{author}{\bibfnamefont{M.~E.~J.} \bibnamefont{Newman}},
  \bibinfo{journal}{Phys. Rev. E} \textbf{\bibinfo{volume}{69}},
  \bibinfo{pages}{066133} (\bibinfo{year}{2004}{\natexlab{c}}).

\bibitem[{\citenamefont{Clauset et~al.}(2004)\citenamefont{Clauset, Newman, and
  Moore}}]{Clauset-Newman-Moore-2004-PRE}
\bibinfo{author}{\bibfnamefont{A.}~\bibnamefont{Clauset}},
  \bibinfo{author}{\bibfnamefont{M.~E.~J.} \bibnamefont{Newman}},
  \bibnamefont{and} \bibinfo{author}{\bibfnamefont{C.}~\bibnamefont{Moore}},
  \bibinfo{journal}{Phys. Rev. E} \textbf{\bibinfo{volume}{70}},
  \bibinfo{pages}{066111} (\bibinfo{year}{2004}).

\bibitem[{\citenamefont{Kernighan and Lin}(1970)}]{Kernighan-Lin-1970-BSTJ}
\bibinfo{author}{\bibfnamefont{B.~W.} \bibnamefont{Kernighan}}
  \bibnamefont{and} \bibinfo{author}{\bibfnamefont{S.}~\bibnamefont{Lin}},
  \bibinfo{journal}{Bell System Technical Journal}
  \textbf{\bibinfo{volume}{49}}, \bibinfo{pages}{291} (\bibinfo{year}{1970}).

\bibitem[{\citenamefont{Capocci et~al.}(2004)\citenamefont{Capocci, Servedio,
  Caldarelli, and Colaiori}}]{Capocci-Servedio-Caldarelli-Colaiori-2004-LNCS}
\bibinfo{author}{\bibfnamefont{A.}~\bibnamefont{Capocci}},
  \bibinfo{author}{\bibfnamefont{V.~D.~P.} \bibnamefont{Servedio}},
  \bibinfo{author}{\bibfnamefont{G.}~\bibnamefont{Caldarelli}},
  \bibnamefont{and} \bibinfo{author}{\bibfnamefont{F.}~\bibnamefont{Colaiori}},
  \bibinfo{journal}{Lecture Notes in Computer Science}
  \textbf{\bibinfo{volume}{3243}}, \bibinfo{pages}{181} (\bibinfo{year}{2004}).

\bibitem[{\citenamefont{Donetti and Mu{\~n}oz}(2005)}]{Donetti-Munoz-2005}
\bibinfo{author}{\bibfnamefont{L.}~\bibnamefont{Donetti}} \bibnamefont{and}
  \bibinfo{author}{\bibfnamefont{M.~A.} \bibnamefont{Mu{\~n}oz}}, in
  \emph{\bibinfo{booktitle}{Proceedings of the 8th Granada Seminar -
  Computational and Statistical Physics}} (\bibinfo{year}{2005}).

\bibitem[{\citenamefont{Wu and Huberman}(2004)}]{Wu-Huberman-2004-EPJB}
\bibinfo{author}{\bibfnamefont{F.}~\bibnamefont{Wu}} \bibnamefont{and}
  \bibinfo{author}{\bibfnamefont{B.~A.} \bibnamefont{Huberman}},
  \bibinfo{journal}{Eur. Phys. J. B} \textbf{\bibinfo{volume}{38}},
  \bibinfo{pages}{331} (\bibinfo{year}{2004}).

\bibitem[{\citenamefont{Fortunato et~al.}(2004)\citenamefont{Fortunato, Latora,
  and Marchiori}}]{Fortunato-Latora-Marchiori-2004-PRE}
\bibinfo{author}{\bibfnamefont{S.}~\bibnamefont{Fortunato}},
  \bibinfo{author}{\bibfnamefont{V.}~\bibnamefont{Latora}}, \bibnamefont{and}
  \bibinfo{author}{\bibfnamefont{M.}~\bibnamefont{Marchiori}},
  \bibinfo{journal}{Phys. Rev. E} \textbf{\bibinfo{volume}{70}},
  \bibinfo{pages}{056104} (\bibinfo{year}{2004}).

\bibitem[{\citenamefont{Reichardt and
  Bornholdt}(2004)}]{Reichardt-Bornholdt-2004-PRL}
\bibinfo{author}{\bibfnamefont{J.}~\bibnamefont{Reichardt}} \bibnamefont{and}
  \bibinfo{author}{\bibfnamefont{S.}~\bibnamefont{Bornholdt}},
  \bibinfo{journal}{Phys. Rev. Lett.} \textbf{\bibinfo{volume}{93}},
  \bibinfo{pages}{218701} (\bibinfo{year}{2004}).

\bibitem[{\citenamefont{Wilkinson and
  Huberman}(2004)}]{Wilkinson-Huberman-2004-PNAS}
\bibinfo{author}{\bibfnamefont{D.~M.} \bibnamefont{Wilkinson}}
  \bibnamefont{and} \bibinfo{author}{\bibfnamefont{B.~A.}
  \bibnamefont{Huberman}}, \bibinfo{journal}{Proc. Natl. Acad. Sci. USA}
  \textbf{\bibinfo{volume}{101}}, \bibinfo{pages}{5241} (\bibinfo{year}{2004}).

\bibitem[{\citenamefont{Young et~al.}()\citenamefont{Young, Sager, Cs{\'a}rdi,
  and H{\'a}ga}}]{Young-Sager-Csardi-Haga-2004-XXX}
\bibinfo{author}{\bibfnamefont{M.}~\bibnamefont{Young}},
  \bibinfo{author}{\bibfnamefont{J.}~\bibnamefont{Sager}},
  \bibinfo{author}{\bibfnamefont{G.}~\bibnamefont{Cs{\'a}rdi}},
  \bibnamefont{and} \bibinfo{author}{\bibfnamefont{P.}~\bibnamefont{H{\'a}ga}},
  \bibinfo{note}{cond-mat/0408263}.

\bibitem[{\citenamefont{Bagrow and Bollt}()}]{Bagrow-Bollt-2005-XXX}
\bibinfo{author}{\bibfnamefont{J.~P.} \bibnamefont{Bagrow}} \bibnamefont{and}
  \bibinfo{author}{\bibfnamefont{E.~M.} \bibnamefont{Bollt}},
  \bibinfo{note}{cond-mat/0412482}.

\bibitem[{\citenamefont{Son et~al.}()\citenamefont{Son, Jeong, and
  Noh}}]{Son-Jeong-Noh-2005-XXX}
\bibinfo{author}{\bibfnamefont{S.-W.} \bibnamefont{Son}},
  \bibinfo{author}{\bibfnamefont{H.}~\bibnamefont{Jeong}}, \bibnamefont{and}
  \bibinfo{author}{\bibfnamefont{J.~D.} \bibnamefont{Noh}},
  \bibinfo{note}{cond-mat/0502672}.

\bibitem[{\citenamefont{de~Nooy et~al.}(2005)\citenamefont{de~Nooy, Mrvar, and
  Batagelj}}]{deNooy-Mrvar-Batagelj-2005}
\bibinfo{author}{\bibfnamefont{W.}~\bibnamefont{de~Nooy}},
  \bibinfo{author}{\bibfnamefont{A.}~\bibnamefont{Mrvar}}, \bibnamefont{and}
  \bibinfo{author}{\bibfnamefont{V.}~\bibnamefont{Batagelj}},
  \emph{\bibinfo{title}{Exploratory Social Network Analysis with Pajek}}
  (\bibinfo{publisher}{Cambridge University Press},
  \bibinfo{address}{Cambridge}, \bibinfo{year}{2005}).

\bibitem[{\citenamefont{West et~al.}(1997)\citenamefont{West, Brown, and
  Enquist}}]{West-Brown-Enquist-1997-Science}
\bibinfo{author}{\bibfnamefont{G.~B.} \bibnamefont{West}},
  \bibinfo{author}{\bibfnamefont{J.~H.} \bibnamefont{Brown}}, \bibnamefont{and}
  \bibinfo{author}{\bibfnamefont{B.~J.} \bibnamefont{Enquist}},
  \bibinfo{journal}{Science} \textbf{\bibinfo{volume}{276}},
  \bibinfo{pages}{122} (\bibinfo{year}{1997}).

\bibitem[{\citenamefont{Enquist et~al.}(1998)\citenamefont{Enquist, Brown, and
  West}}]{Enquist-Brown-West-1998-Nature}
\bibinfo{author}{\bibfnamefont{B.~J.} \bibnamefont{Enquist}},
  \bibinfo{author}{\bibfnamefont{J.~H.} \bibnamefont{Brown}}, \bibnamefont{and}
  \bibinfo{author}{\bibfnamefont{G.~B.} \bibnamefont{West}},
  \bibinfo{journal}{Nature} \textbf{\bibinfo{volume}{395}},
  \bibinfo{pages}{163} (\bibinfo{year}{1998}).

\bibitem[{\citenamefont{West et~al.}(1999)\citenamefont{West, Brown, and
  Enquist}}]{West-Brown-Enquist-1999-Science}
\bibinfo{author}{\bibfnamefont{G.~B.} \bibnamefont{West}},
  \bibinfo{author}{\bibfnamefont{J.~H.} \bibnamefont{Brown}}, \bibnamefont{and}
  \bibinfo{author}{\bibfnamefont{B.~J.} \bibnamefont{Enquist}},
  \bibinfo{journal}{Science} \textbf{\bibinfo{volume}{284}},
  \bibinfo{pages}{1677} (\bibinfo{year}{1999}).

\bibitem[{\citenamefont{Enquist et~al.}(1999)\citenamefont{Enquist, West,
  Charnov, and Brown}}]{Enquist-West-Charnov-Brown-1999-Nature}
\bibinfo{author}{\bibfnamefont{B.~J.} \bibnamefont{Enquist}},
  \bibinfo{author}{\bibfnamefont{G.~B.} \bibnamefont{West}},
  \bibinfo{author}{\bibfnamefont{E.~L.} \bibnamefont{Charnov}},
  \bibnamefont{and} \bibinfo{author}{\bibfnamefont{J.~H.} \bibnamefont{Brown}},
  \bibinfo{journal}{Nature} \textbf{\bibinfo{volume}{401}},
  \bibinfo{pages}{907} (\bibinfo{year}{1999}).

\bibitem[{\citenamefont{Banavar et~al.}(1999)\citenamefont{Banavar, Maritan,
  and Rinaldo}}]{Banavar-Maritan-Rinaldo-1999-Nature}
\bibinfo{author}{\bibfnamefont{J.~R.} \bibnamefont{Banavar}},
  \bibinfo{author}{\bibfnamefont{A.}~\bibnamefont{Maritan}}, \bibnamefont{and}
  \bibinfo{author}{\bibfnamefont{A.}~\bibnamefont{Rinaldo}},
  \bibinfo{journal}{Nature} \textbf{\bibinfo{volume}{399}},
  \bibinfo{pages}{130} (\bibinfo{year}{1999}).

\bibitem[{\citenamefont{Enquist et~al.}(2003)\citenamefont{Enquist, Economo,
  Huxman, Allen, Ignace, and
  Gillooly}}]{Enquist-Economo-Huxman-Allen-Ignace-Gillooly-2003-Nature}
\bibinfo{author}{\bibfnamefont{B.~J.} \bibnamefont{Enquist}},
  \bibinfo{author}{\bibfnamefont{E.~P.} \bibnamefont{Economo}},
  \bibinfo{author}{\bibfnamefont{T.~E.} \bibnamefont{Huxman}},
  \bibinfo{author}{\bibfnamefont{A.~P.} \bibnamefont{Allen}},
  \bibinfo{author}{\bibfnamefont{D.~D.} \bibnamefont{Ignace}},
  \bibnamefont{and} \bibinfo{author}{\bibfnamefont{J.~F.}
  \bibnamefont{Gillooly}}, \bibinfo{journal}{Nature}
  \textbf{\bibinfo{volume}{423}}, \bibinfo{pages}{639} (\bibinfo{year}{2003}).

\bibitem[{\citenamefont{Garlaschelli et~al.}(2003)\citenamefont{Garlaschelli,
  Caldarelli, and Pietronero}}]{Garlaschelli-Caldarelli-Pietronero-2003-Nature}
\bibinfo{author}{\bibfnamefont{D.}~\bibnamefont{Garlaschelli}},
  \bibinfo{author}{\bibfnamefont{G.}~\bibnamefont{Caldarelli}},
  \bibnamefont{and}
  \bibinfo{author}{\bibfnamefont{L.}~\bibnamefont{Pietronero}},
  \bibinfo{journal}{Nature} \textbf{\bibinfo{volume}{423}},
  \bibinfo{pages}{165} (\bibinfo{year}{2003}).

\end{thebibliography}

\clearpage

\begin{itemize}
  \item Figure 1:Ammonia plant process flow diagram with the major
equipment. 1~--~gasification reactor, 2~--~carbon scrubber,
3~--~H$_2$S absorber, 4~--~humidifier, 5~--~shift converter,
6~--~dehumidifier, 7~--~CO$_2$ absorber, 8~--~air compressor,
9~--~pressure column, 10~--~nitrogen compressor, 11~--~nitrogen wash
column, 12~--~synthesis gas compressor, 13~--~ammonia converter,
14~--~unitized chiller, 15~--~liquid ammonia tank.
  \item Figure 2: Rank-ordering analysis of the in-degree, out-degree, and
all-degree of the AP network. We have translated vertically the
in-degree line by 4 and the out-degree line by 25 for better
presentation. The lines are the best fit of tail distribution to
(\ref{Eq:xR}).
  \item Figure 3: (Color online) Modular structure of the AP network. The
shapes of the vertices represent different sections of the process
flow of the Ammonia Plant. This figure was produced with Pajek
\cite{deNooy-Mrvar-Batagelj-2005}.
  \item Figure 4: Power-law scaling of $S$ against $A$. The line represents
the power-law fit to the data.
\end{itemize}

\clearpage

\clearpage

\clearpage

\clearpage

\end{document}